\documentclass[fleqn,usenatbib]{mnras}

\usepackage{newtxtext,newtxmath}

\usepackage[T1]{fontenc}

\DeclareRobustCommand{\VAN}[3]{#2}
\let\VANthebibliography\thebibliography
\def\thebibliography{\DeclareRobustCommand{\VAN}[3]{##3}\VANthebibliography}

\usepackage{graphicx}	
\usepackage{amsmath}	

\title[Search for Dyson Rings Around Pulsars]{Search for Dyson Rings Around Pulsars: Unexpected Light Curves}

\author[O. Kayali et al.]{
Ogetay Kayali,$^{1}$\thanks{E-mail: okayali@mtu.edu}
Emir Haliki,$^{2}$
Kubra Bas$^{3}$
and Robert J. Nemiroff$^{1}$
\\
$^{1}$Michigan Technological University\\1400 Townsend Drive\\Houghton, MI 49931,USA\\
$^{2}$Ege University\\Izmir, Turkey\\
$^{3}$Technische Universität München\\München, Germany
}

\date{Accepted XXX. Received YYY; in original form ZZZ}

\pubyear{\the\year{}}

\begin{document}
\label{firstpage}
\pagerange{\pageref{firstpage}--\pageref{lastpage}}
\maketitle

\begin{abstract}
Finding Dyson rings around distant pulsars may involve identifying light curve features that have not been previously identified. Previous studies covered the detection of a ring structure uniformly brightened by the central pulsar, mostly in infrared light. Here, more complex light curves are explored, which arise inherently from the pulsar beam spot's commonly predicted superluminal speed. These speeds may cause multiple images of the pulsar's spot on the Dyson ring to appear simultaneously to a distant observer, and so feature bright creation and annihilation events. Therefore, it is possible that even if Dyson ring structures had been observed previously, they might have remained unnoticed. Similar light curve features may appear on naturally occurring dust rings around pulsars that reflect detectable pulsar radiation. 
\end{abstract}




\section{Introduction}

The Kardashev scale \citep{Kardashev_1964} characterizes the technological levels of extraterrestrial civilizations, where a civilization's energy consumption defines that civilization's level. Humanity, on this scale, is currently Type I, which indicates a power usage of the order of $4\times 10^{19}\text{ erg s}^{-1}$. A civilization that can harvest total power in order ($4\times 10^{33}\text{ erg s}^{-1}$) is considered Type II. \citet{Dyson_1960} suggested that advanced civilizations of Type II or higher might construct a spherical shell with a radius of about 1 AU that surrounds their host star to harvest its energy. However, \cite{DeBiase_2008, Shaw_2012, Osmanov_2019} showed that a spherical shell is not ideal. 

Full spheres at 1 AU and farther require enormous amounts of material, possibly more than what exists in the entire star system. Therefore, the concept of the Dyson Sphere has been extended to different structures known as Dyson swarms and Dyson rings \citep{Harp_2016, Osmanov_2018, Haliki_2020, Shostak_2020, Wright_2020, Hsiao_2021}. For pulsars, power can be usefully harvested using only a small fraction of a full Dyson sphere that tracks the pulsar's beam. Isolating the pulsar's beam track naturally results in a ring -- which could be engineered to coincide with a Dyson ring \citep{osmanov_2016}.

\begin{figure}
    \includegraphics[width=9cm]{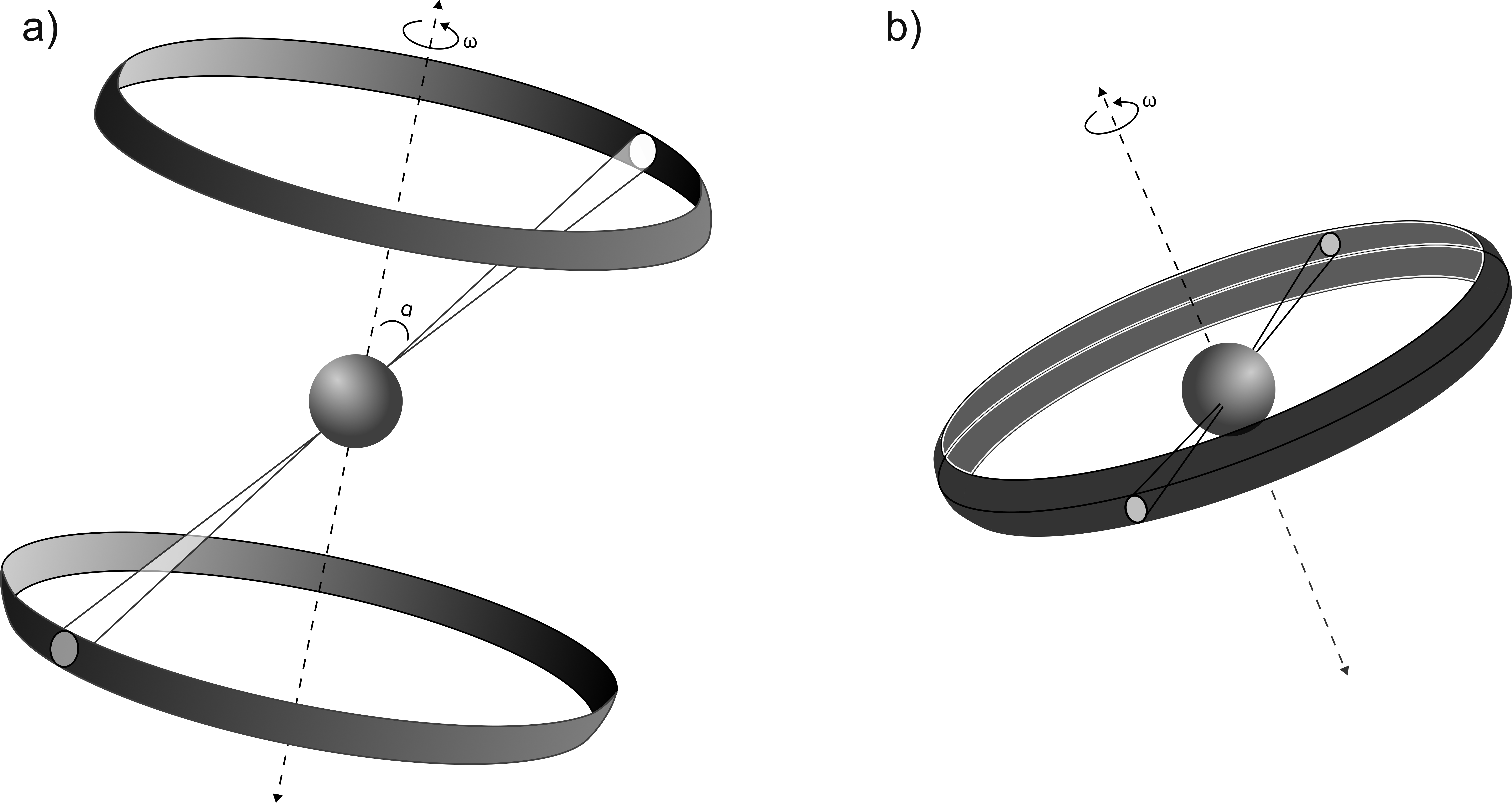}
    \centering
    \caption{Two ring-like Dyson structures are shown. In subfigure a, the inclination angle of pulsar $\alpha$ is not close to $90^\circ$ and in order to gather the energy from both directions, two separate ring structures are needed. In subfigure b, the inclination angle $\alpha$ is close to $90^\circ$ so that one larger ring can intersect both beams.}
    \label{fig:ring_structure}
\end{figure}

Depending on the magnetic inclination angle $\alpha$ of the pulsar's beam with the pulsar's rotation axis, the Dyson ring can be a single ring-like cross-section of a sphere. For smaller $\alpha$, it can be two parallel ring structures as shown in Figure \ref{fig:ring_structure}. However, the latter is predicted to be unstable as discussed in Section \ref{sec:DysonRings}. Henceforth, in this paper, a single-ring structure will be assumed. Due to the powerful ($ \sim 10^{29}\text{ erg s}^{-1}$) nature of the beam \citep{Bagchi_2013, Szary_2014}, depending on the period of the pulsar, this ring structure must be larger than a critical minimum radius \citep{osmanov_2016} or either the beam will not be able to form, or it will destroy the Dyson ring. Also, due to the finite amount of planetary materials available, the ring structure must be smaller than a critical maximum radius. 

\begin{figure}
    \includegraphics[width=9cm]{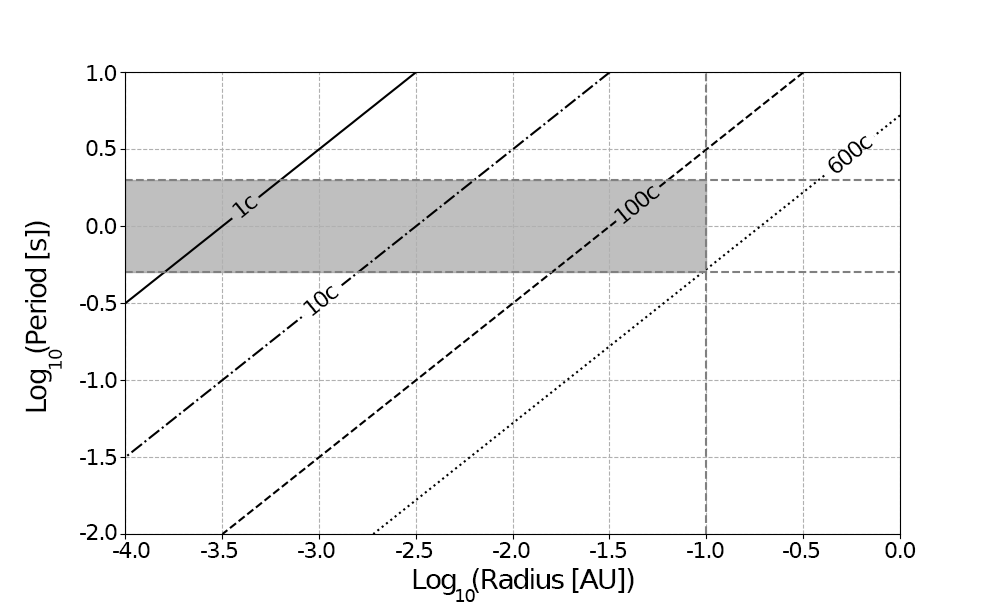}
    \centering
    \caption{Beam speeds on the surface of the ring as a function of period $p$ and radius $R$. The shaded area indicates the regime defined by \citep{Osmanov_2018}. The solid line, dotted dashed line, dashed line, and dotted line indicate beam spot speeds of $c$, $10c$, $100c$, and $600c$ respectively.}
    \label{fig:beam_speeds}
\end{figure}

Considering the minimum radius constraint, it is readily seen that in many scenarios the sweeping speed of the beam spot is faster than light, as shown in Figure \ref{fig:beam_speeds}. In such cases, a little-known phenomenon called Relativistic Image Doubling (RID) takes place \citep{Nemiroff_2023}. This results in a bright flash followed by an observer-perceived creation of two images of the beam spot that move in opposite directions. These two beam-spot images have varying angular speeds that lead to a non-uniform perceived brightening of the ring. 

Since the non-uniform brightening significantly impacts the possible detection of Dyson rings, understanding the nature of this brightening becomes important. Characteristic light curves of the pulsar-brightened ring depend on the sweeping beam spot's speed $v$ and the inclination angle $\theta$ of the plane of the ring to the plane of the sky. Note that, outside of perception, the pulsar only creates one beam spot on the Dyson ring, per pole. The speed $v$ of this spot depends on the ring's radius $R$ and the pulsar's period $p$. Therefore, different period and radius configurations might result in similar sweeping speeds.  

A pulsar's beam may create light across a wide range of the electromagnetic spectrum, as pulsars are known to emit radiation from the radio to the gamma-ray. Therefore, this work is generalizable to all pulsar radiations that either reflect off a ring surface or heat a ring structure in a way that causes subsequent emission. Therefore, this work demonstrates that, in many cases, analyses that ignore observable RID-related brightening effects may arrive at incorrect conclusions. In addition, the theory of RID on rings is not limited to Dyson rings. For example, naturally occurring dust rings around pulsars may absorb or reflect detectable radiation in a similar manner.

In this paper, we investigate how light curve profiles differ from the classically expected cases for ring structures. Section \ref{sec:RID} focuses on the concepts behind RID relevant to Dyson rings around pulsars and presents useful formulae. Section \ref{sec:DysonRings} presents the possible ring-like structures and reviews the geometry and the stability of such geometries. Then, based on these geometries, Section \ref{sec:LightCurves} provides new light curves. Section \ref{sec:Discussion} concludes with some discussion.


\section{Relativistic Image Doubling (RID)} \label{sec:RID}

In the late nineteenth century, \citet{Rayleigh_1896} predicted that if supersonic sounds were heard by a stationary observer, these sounds could be heard time reversed: backwards. However, he did not predict a pair event where two sound "images" would be heard simultaneously forward and backward. To the best of our knowledge, the earliest mention of the pair event phenomenon was over 50 years ago by \citet{Cavaliere_1971} who briefly hypothesized that such a pair event might explain apparent superluminal motion in quasars. \citet{Bolotovskii_1990} showed that when a charged particle moving along a circular trajectory emits radiation, this leads to an observation of image creation and image annihilation events. About 15 years ago, the apparent "splitting" of a superluminal spot over a hypothetical wall by a rotating light source was shown by \citet{Baune_2009}. About 10 years ago, Relativistic Image Doubling (RID) was analyzed in detail by \citet{Nemiroff_2015}. Shortly thereafter, a pair creation event was first measured in the lab for plane waves incident on a tilted screen by \citet{Clerici_2016}. 

Relativistic image doubling (RID) occurs both in objective reality, outside the view of any one observer, and in the subjective observations of a single observer. For a single observer, RID is seen to occur when the illumination spot's speed toward the observer drops from superluminal to subluminal \citep{Nemiroff_2015, Nemiroff_2023}. In this context, the term "spot" can be generalized to any number of intangible phenomena, such as a laser's spot on a wall, a shadow, and even tangible entities, such as a muon that moves faster than light does through water. The fundamental reason behind the high relative brightness of observer-oriented RID is that the spot itself moves near to -- and above -- the speed of its own emissions. When this happens, an observer sees a source across many locations in a short time, causing a peak in the source's apparent brightness, even if the source is emitting uniformly and isotropically. Stated differently, new photons that are emitted at a closer position arrive nearly simultaneously with old photons that were already emitted at a more distant position \citep{Nemiroff_2015, Nemiroff_2018}. 

\begin{figure}
    \includegraphics[width=9cm]{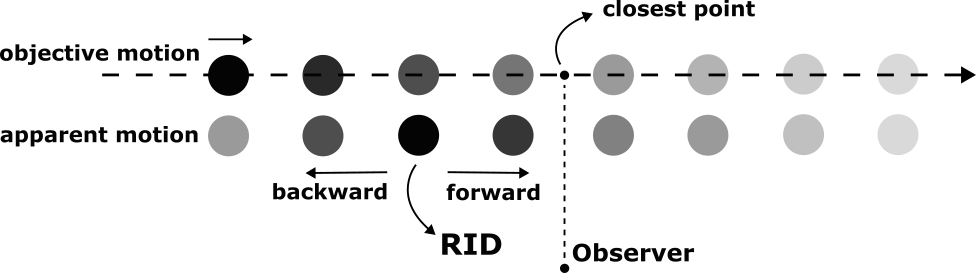}
    \centering
    \caption{Observed Relativistic Image Doubling (RID) is diagrammed for a straight spot trajectory, where the spot moves left to right faster than light and the observer is to the right of the perceived RID location. The top row represents objective reality, while the bottom row represents how RID appears from the perspective of the observer. The shading represents the time frame, and black and white denote the initial and final positions, respectively. After RID occurs, two perceived images of the spot are created and diverge.}
    \label{fig:RID_straight}
\end{figure}

In the case of RID, the spot's first observed position is not the closest point on its track to the observer, but rather somewhere farther away. RID starts with the apparent creation of two spots as, formally, an infinitely bright flash. Subsequently, the two observed spots diverge, one going forward in the direction that the spot is actually moving, and the other going backward \citep{Nemiroff_2018}. These two images move with different angular speeds and brightness over time. Observed RID onset is here called an image creation (IC) event. The appearance of RID for a straight trajectory is shown in Figure \ref{fig:RID_straight}. 

For physical objects, RID can be detected by observing the Cherenkov light where the charged particle, such as a muon, moves faster than the speed of light of the medium \citep{Kaushal_2020, Kayali_2022, Nemiroff_2020}. Physical objects are, of course, limited to move less than $c$, the speed of light in vacuum. Conversely, for non-physical objects such as beam spots, there is no speed limit. Special Relativity (SR) is not violated in either case. 

In objective reality, in cases where the rotational period of a pulsar is relatively short, and its surrounding Dyson ring is relatively large, the pulsar's beam may sweep across the surface of the ring faster than light, as shown in Figure \ref{fig:beam_speeds}. Due to the circular symmetry, image creation (IC) is followed by image annihilation (IA) where two images merge into a formally infinite flash and then disappear. However, the formation of IC and IA strictly depends on the inclination angle $\theta$ of the Dyson ring to the observer. In face-on cases ($\theta=\pi/2$), the beam sweeps across the entire structure with the same radial velocity from the observer's angle of view, hence no RID event will appear. However, cases where some parts of the Dyson ring are farther from the observer than other parts, RID might be observed. For the extreme edge-on case ($\theta=0$), due to the circular shape of the structure, as the beam moves towards the nearest and farthest points, the apparent radial speed drops to zero. If the radial speed of the spot is ever superluminal on the Dyson ring's surface, this guarantees IC and IA formation. 

\begin{figure}
    \includegraphics[width=9cm]{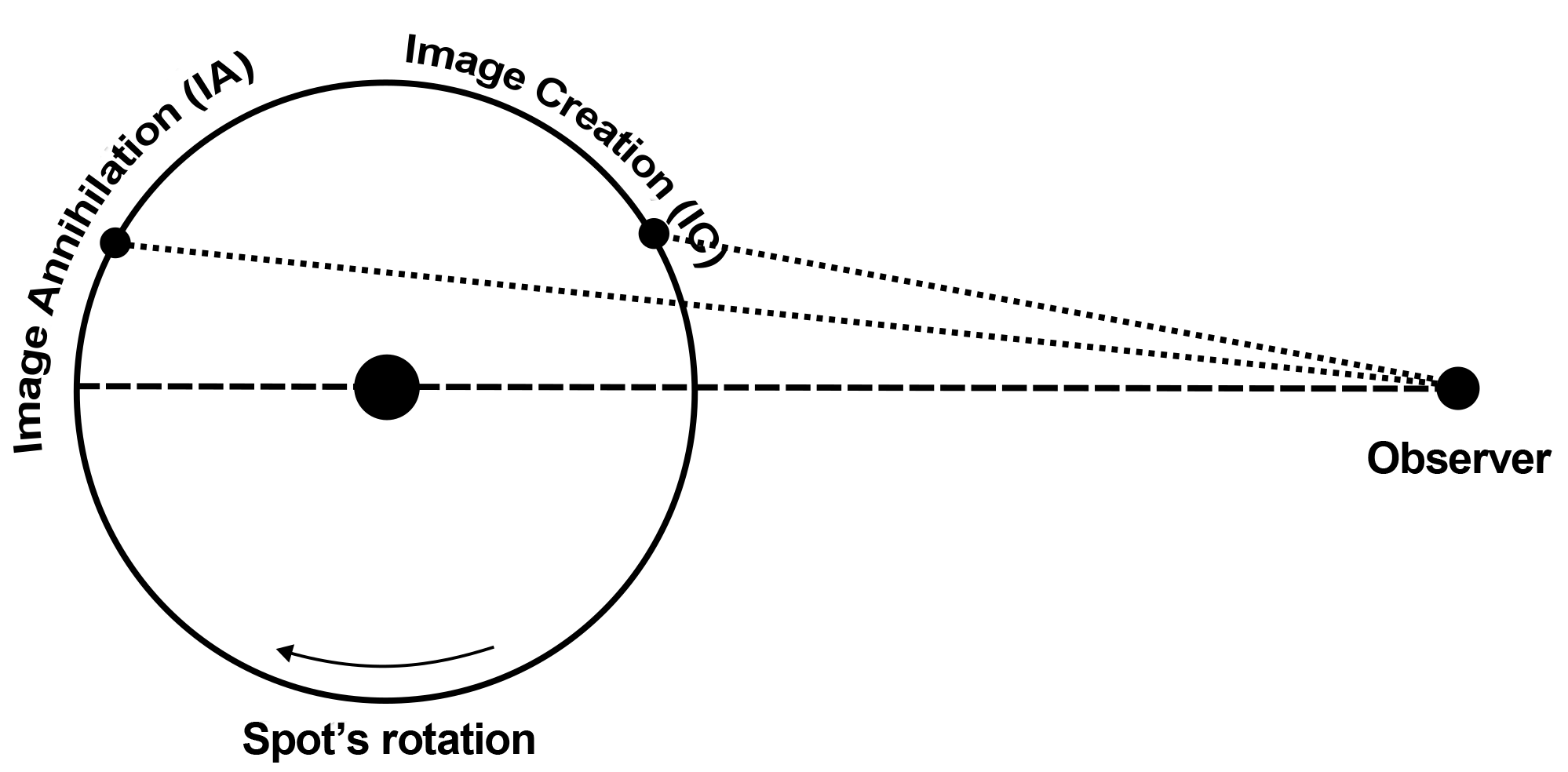}
    \centering
    \caption{An image creation (IC) and image annihilation (IA) events around a circle where the central pulsar sweeps the circle clockwise.}
    \label{fig:RID_around_circle}
\end{figure}

Since IC and IA are symmetric with each other for distant observers, as shown in Figure \ref{fig:RID_around_circle}, to observe both IC and IA, the spot must be observable over the entire ring, not just on the more distant half. Therefore, if the ring is opaque, the reflected radiation only provides one event. However, thermal radiation can resolve both events, although the temperature gradient that is created by the sweeping beam needs to be well-known to determine the light curve. Of course, the temperature differences might be too small to be observable, but since in this paper such a hypothetical structure's material is only limited by our imagination, we discuss the cases where the observation is possible.

In the case of IC or IA, the two images move at different speeds along their trajectories. Different IC and IA events occur at different times and have the same theoretical maximum brightness. For Dyson rings around pulsars, the beam spot's speed can be relatively high compared to the speed of light as shown in Figure \ref{fig:beam_speeds}, and images can persist from previous cycles, therefore yielding multiple IC and IA events and multiple spot images visible on the ring simultaneously. As RID occurs as a very bright flash, the light curve appears to be peaking at different times depending on the speed of the beam. Therefore, the light curves of such structures significantly differ from what is expected in classical cases. The number of spot images that appear at any one time can be calculated from the parameters $p$, $\theta$, and $R$, where $p$ is the period of the pulsar, $\theta$ is the inclination angle of the ring, and $R$ is the ring's radius. 

\subsection{Math}\label{sec:MathRID}

The Dyson ring radius optimized to harvest the most power of the central pulsar depends, in part, on the brightness of the pulsar beam which correlates, in turn, with the rotational period of that pulsar \citep{osmanov_2016}. Additional assumptions include that the Dyson ring is far enough out that the pulsar's beam does not destroy the ring, that the beam has enough room to fully form, and that the Dyson ring is in the habitable zone of the pulsar \citep{osmanov_2016}. Pulsars exhibit a wide range of rotational periods, from milliseconds to seconds. At the fast rotation end, Dyson rings around millisecond pulsars with periods $p=(0.01-0.05)$s have been found to be practical only when their radii are quite large \citep{Osmanov_2018}. This leads to a Dyson ring mass that likely exceeds the total planetary mass by several orders of magnitude. However, for relatively slow rotating pulsars with $p = (0.5-2)$ s, practical and optimal Dyson ring radii $R$ do exist and are predicted to be in the range $R = (10^{-4}-10^{-1})$ AU. Even for these slowly rotating pulsars, though, the speed of the sweeping pulsar beam on the Dyson ring can be faster than light, as shown in Figure \ref{fig:beam_speeds}.

\begin{figure}
    \includegraphics[height=11cm]{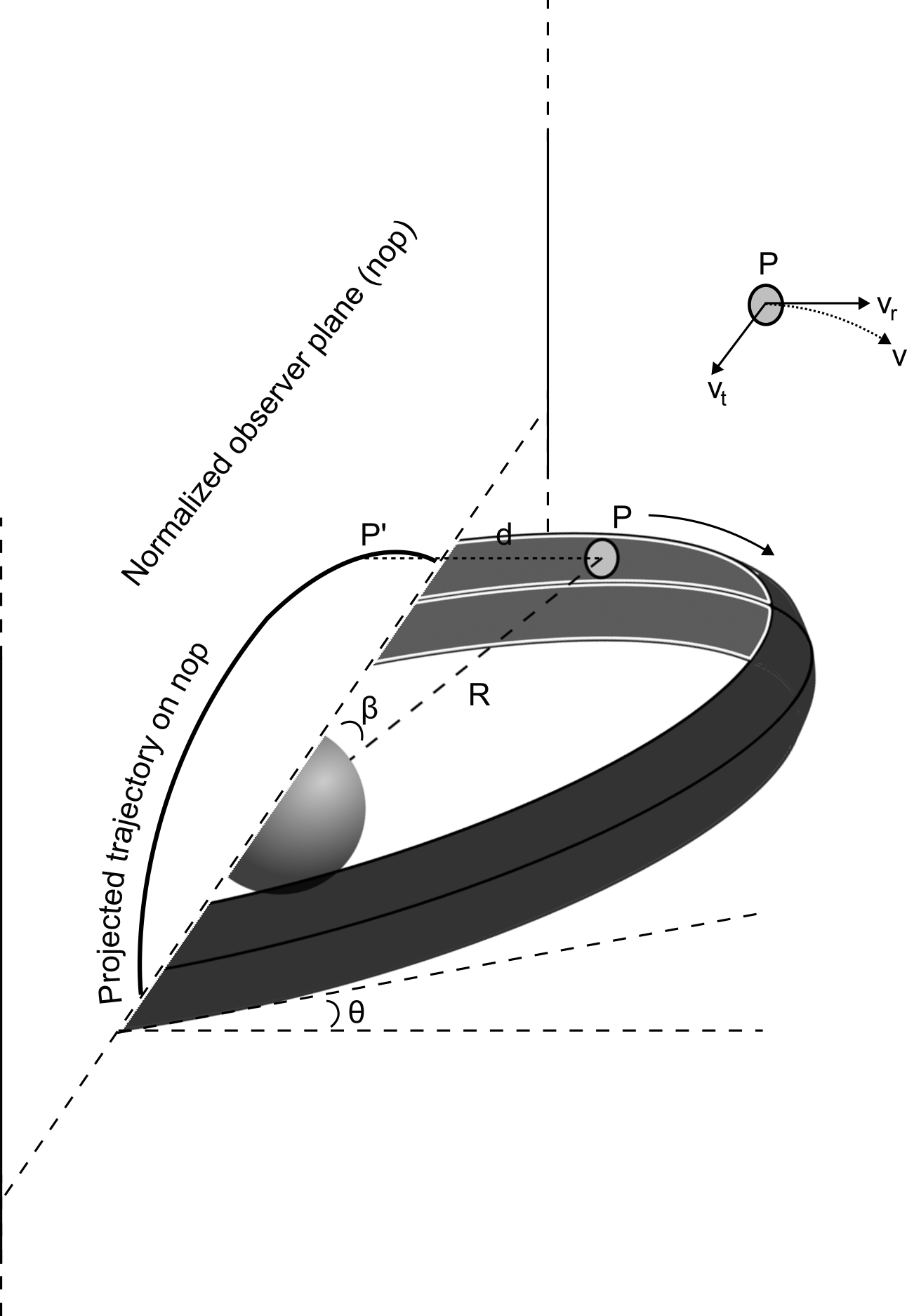}
    \centering
    \caption{The geometry of RID over a Dyson ring around a pulsar where the observer is infinitely far on the left so that the projected trajectory represents the angular motion of the spot's trajectory from the observer's point of view. Here $\beta$ and $\theta$ define the angular position of the spot over the ring and the inclination angle of the ring from the observer's point of view, respectively. $R$ and $d$ are the radius of the ring and the truncated distance to the observer, respectively, as explained in Section \ref{sec:MathRID}.}
    \label{fig:ring_projection}
\end{figure}

To find when RID occurs, the relative time of observation of a beam spot at a particular point $P$ needs to be calculated, which is given by $t_\text{tot} = t_\text{rot} + t_\text{beam} + t_\text{ref}$. Here $t_\text{rot}$ denotes the time for the pulsar's axis to rotate to that particular point, $t_\text{beam}$ is the time taken by the beam to reach the ring's surface, and $t_\text{ref}$ is the time taken by the reflected light to reach the observer from the surface of the ring. If one considers thermal radiation, another time constant that comes from the cooling timescale of the element should be added. In an expanded form 
\begin{equation}
    \centering
    \label{eq:ttot}
    t_\text{tot}= \Bigg(p\frac{\beta}{2\pi} \Bigg) + \frac{R}{c} + \frac{1}{c} \Bigg[R\cos{\theta} (1+\sin{\beta})   \Bigg] ,
\end{equation}

\noindent
where $\beta$ is the angular position of the beam on the ring. This geometry is shown in Figure \ref{fig:ring_projection}. To find the angular position of the onset of RID IC or the conclusion of RID IA, one needs to evaluate $\frac{dt_\text{tot}}{d_\beta}=0$. This can be readily seen to yield $\cos\theta=-\frac{Pc}{2\pi R \cos\beta}$ and since $2\pi R/p=v$ by definition
\begin{equation}
    \centering
    \label{eq:tmin}
    \cos\beta=-\frac{c}{v\cos\theta} .
\end{equation}

\noindent
An IC event is seen to occur when the radial speed of the spot drops from superluminal to subluminal. Of course, this can only happen when the beam spot's speed on the disk is superluminal to begin with. Similarly, an IA event is seen to occur when the radial speed of the spot increases from subluminal to superluminal. But both IC and IA events happen only on one side of the ring where the spot moves toward the observer as shown in Figure \ref{fig:RID_around_circle}.

The case where $v$ goes to infinity can be considered similar to the disk echo of a bright isotropic flash by the central star. In this intuitively simpler case, it can be readily seen that IC occurs at the closest point ($\beta=-\pi/2$) and IA occurs at the furthest point ($\beta=\pi/2$) to the observer. This hypothetical infinitely fast case results in the maximum separation of $180^\circ$ between the IC and IA events and the separation becomes smaller as $v$ approaches $c$, ultimately merging and resulting in no RID. Since Eq. \ref{eq:ttot} has the form of $t=k_1\beta +k_2\sin\beta+k_3$ where $k_3=(R+D)/c$ is a constant, this term is neglected when we truncate in time. More intuitively, it corresponds to the geometry depicted in Figure \ref{fig:ring_projection}. 

\section{Pulsars and Dyson Rings} \label{sec:DysonRings}

Dyson ring structures have been hypothesized to have at least two distinct configurations \citep{Haliki_2019} depending on the inclination angle $\alpha$ of the pulsar as shown in Figure \ref{fig:ring_structure}. If $\alpha$ is close to 90 degrees, then only one solid ring structure needs to be built. However, when $\alpha$ significantly deviates from $90$ degrees, then deploying two rings creates an advantage in power absorption. However, as discussed further in this section, the two-ring configuration has been found unstable, or framed differently, would require a significant amount of power to stabilize. Alternatively, the pulsar's beam spots might move in different planes making the geometry even more complex. Nevertheless, for simplicity in this initial paper, only one solid ring structure is considered.

\begin{figure}
    \includegraphics[width=8.5cm]{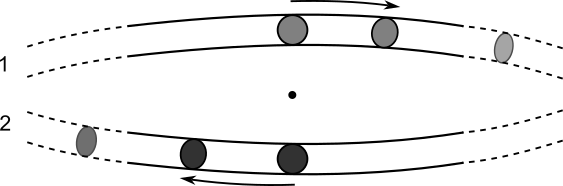}
    \centering
    \caption{A single Dyson ring structure as seen by a distant observer. Here, the rotation of the sweeping beam is clockwise. If only reflected radiation is observed, then only the farther plane is visible. Due to the circular structure, a circular beam spot would become significantly distorted as it moves towards the edges, with its brightness decreased by Lambert's cosine law.}
    \label{fig:ring_alignment}
\end{figure}

In Figure \ref{fig:ring_alignment}, the ring structure is depicted from a distant observer's point of view. The pulsar emits two beams, one from its north magnetic pole, and the other from its south magnetic pole. The pulsar is considered to be in the exact center of the single Dyson ring. Therefore, the beams from both the pulsar's north and south magnetic poles will sweep across the inside of the Dyson ring creating spots that are diametrically opposed. 

If the Dyson structure is opaque, then the beam's reflected spot on the nearest half of the ring will not be visible, which will impact the observed light curve. In these cases, the beam's reflection is only visible from the far part of the Dyson ring. For this reason, in Figure \ref{fig:ring_projection} only the more distant half of the ring structure is pictured. The sweeping beam is assumed, without loss of generality, to move clockwise. The south pole's beam will appear after half the period and have the same characteristics as the north pole's beam. Therefore, the south beam will be a 180-degree phase-shifted copy of the north beam. 

In Figure \ref{fig:ring_projection}, an arbitrary position $P$ represents the location of the north beam's spot on the Dyson ring at time $t$, while $d$, without loss of generality, represents a truncated distance from the spot to a distant observer. In Equation \ref{eq:ttot}, positive values of $\sin \beta$ correspond to the farthest half of the ring, while negative values correspond to the nearest half (not shown). In other words, although the observer is formally at infinity, for didactic and calculational purposes the observer position is considered to be the center of the ring. Because the observer is so far away, it is assumed that all the light rays emerging from the ring structure that are visible to the observer are emitted, effectively, parallel to each other.

Here $\theta$ is the inclination angle of the ring from the observer's point of view and $\theta=0^\circ$ and $\theta=90^\circ$ correspond to edge-on and face-on view cases, respectively. It can be readily seen that the angle $\theta$ affects the physical significance of the events while an angle $\theta^\prime$ that is perpendicular to this angle is not significant but only changes the orientation of the ring in the sky. Therefore, this angle $\theta^\prime$ will be ignored. Using trigonometric identities, the truncated distance $d$ can be found as $d=R\cos(\theta)\sin(\beta)$. Therefore, when $\theta=90^\circ$ (face-on case), $d$ does not depend on the angular position of the beam spot $\beta$ and is always equal to zero. On the other hand, when $\theta=0^\circ$ (edge-on case), $d$ is maximum when $\beta=90^\circ$ (farthest point from the observer) where $d=R$. Consequently, for any $\theta$ and $\beta$ pairs, the distance $d$ must be $-R \leq d \leq R$.

\begin{figure}
    \includegraphics[width=9cm]{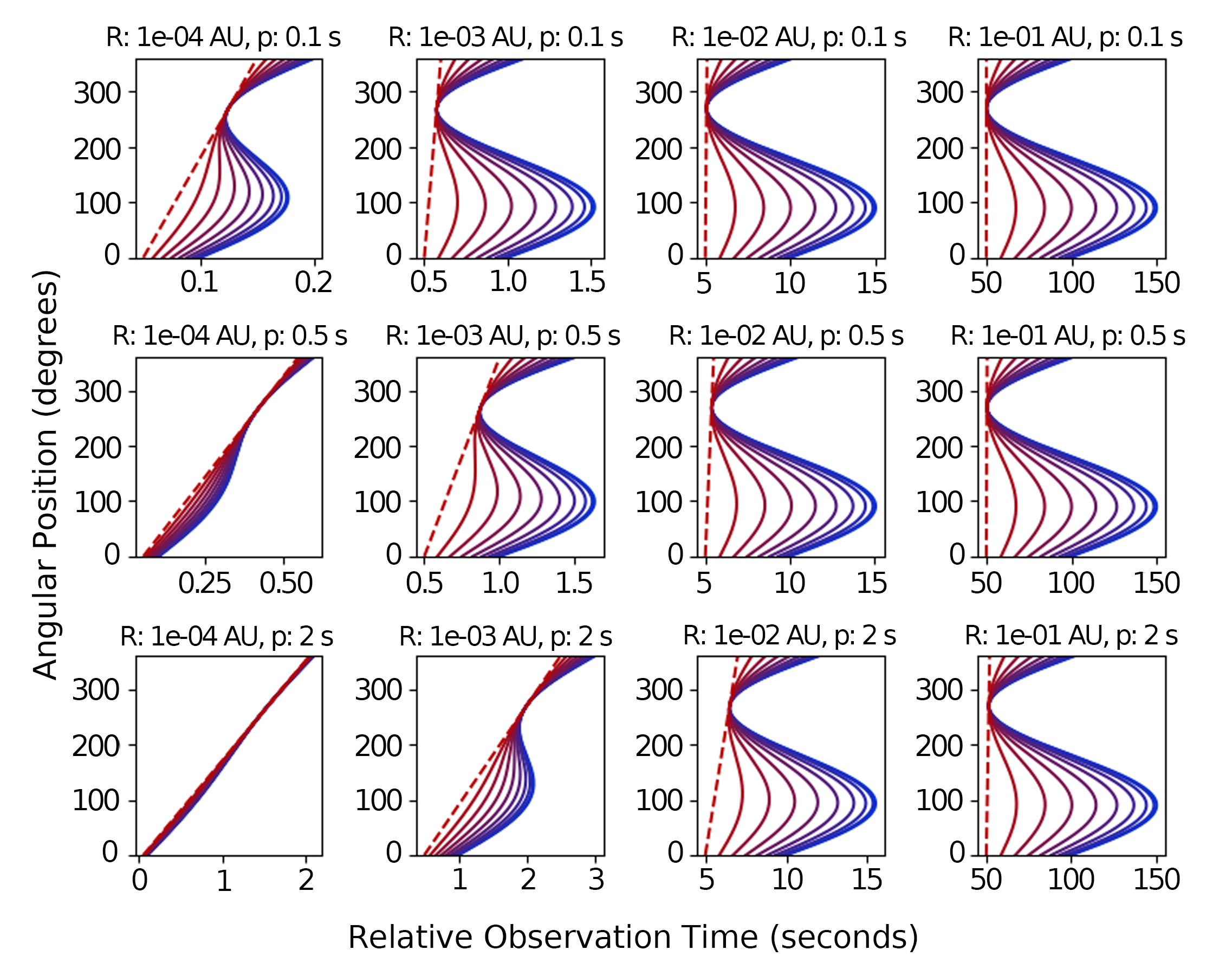}
    \centering
    \caption{Angular position vs. time plots are shown for Dyson ring cases with different $p$, $R$, and $\theta$ values, where relative observation time and angular position of the beam ($\beta$) are represented with the x-axis and y-axis respectively. Here, $\theta$ values are shown with different colors where blue indicates $\theta=0$ (edge-on) and red indicates $\theta=90^\circ$ (face-on) case. }
    \label{fig:various_cases}
\end{figure}

Assuming an advanced civilization builds a Dyson ring in a star's habitable zone, calculations done by \citet{Osmanov_2018} show that for relatively slow rotating pulsars with $P=0.5-2$ s and slow down parameter $\dot{P}=10^{-15}\text{ s s}^{-1}$, the possible ring sizes with a temperature of $T=300-600K$ are in the range of $R=10^{-4}$ AU and $R=10^{-1}$ AU. Figure \ref{fig:beam_speeds} shows this range as a shaded gray area. Inspection of this figure makes it clear that many Dyson ring scales and geometries result in beam spot speeds that are faster than light. In fact, the speeds reach up to $\approx600c$ while $10c$ is approximately the median value. These fast sweeping speeds enable ring configurations with $\theta$ values smaller than $90^\circ$ to exhibit RID as shown in Figure \ref{fig:various_cases}. Technical difficulties of detecting speeds higher than $10c$ are discussed in Section \ref{sec:Discussion}.

It has been shown by \citet{Osmanov_2018} that for millisecond pulsars, the habitable zone lies at $R=10-350$ AU. At these large radii, the amount of rocky material expected within the star system is less than that needed to create even a minimal mass Dyson ring. Here, we do not limit our work to habitable zone assumptions and cover the range between $R=10^{-4}$ AU to $R=1$ AU and $p=10^{-3}$s to $p=10$s. The reasons for this extension are discussed in Section \ref{sec:Discussion}.

There is a distance from the central pulsar past which co-rotation must involve faster than light motion. The radius of this speed of Light Cylinder is defined as $R_{LC} = \frac{c p}{2\pi }$. Although magnetic field lines farther than $R_{LC}$ can be dragged by the central pulsar to superluminal speeds, particles themselves are of course limited to a maximum speed of $c$. For these particles, it is generally considered that the radius of radio emission $r_{em} \le 0.1 R_{LC}$ \citep{Kijak_2003}. If the Dyson ring radius is less than the emission radius such that $R_{Dyson} < r_{em}$, then the pulsar beam would not be fully formed, and so the reflection scenario envisioned here might be much reduced. However, it has been shown that these problems can be avoided by direct particle injection into the pulsar's magnetosphere \citep{Chennamangalam_2015}.

Although such technologies have yet to be developed by humankind, it is possible to critique, in a general sense, the advantages and disadvantages of different architectures for Dyson ring megastructures. As $\alpha$ diverges from $90^\circ$, the amount of material needed to construct a large cross-section of a sphere increases to exceed the amount of material available in the entire star system. To avoid this problem, a two-ring configuration as shown in Figure \ref{fig:ring_structure} might be preferred. However, calculations show that for $\alpha=45^\circ$, the force over each ring is in order of $10^{18}$ Newtons and the amount of power required to keep the rings stable might exceed $10^{33}$ erg s$^{-1}$ \citep{Haliki_2019}. Although the required power could be obtained from the pulsar's beam with a radius $R = R_{LC}/10$, gravitationally non-symmetric geometries would have high perturbations that cause oscillations that would cause the ring structure to become unstable. Because of that, it is generally assumed that a two-ring Dyson configuration is unworkable and henceforth, only a one-ring configuration will be considered.


\section{Dyson Ring Light Curves with Relativistic Image Doubling} \label{sec:LightCurves}

\begin{figure}
    \includegraphics[width=9cm]{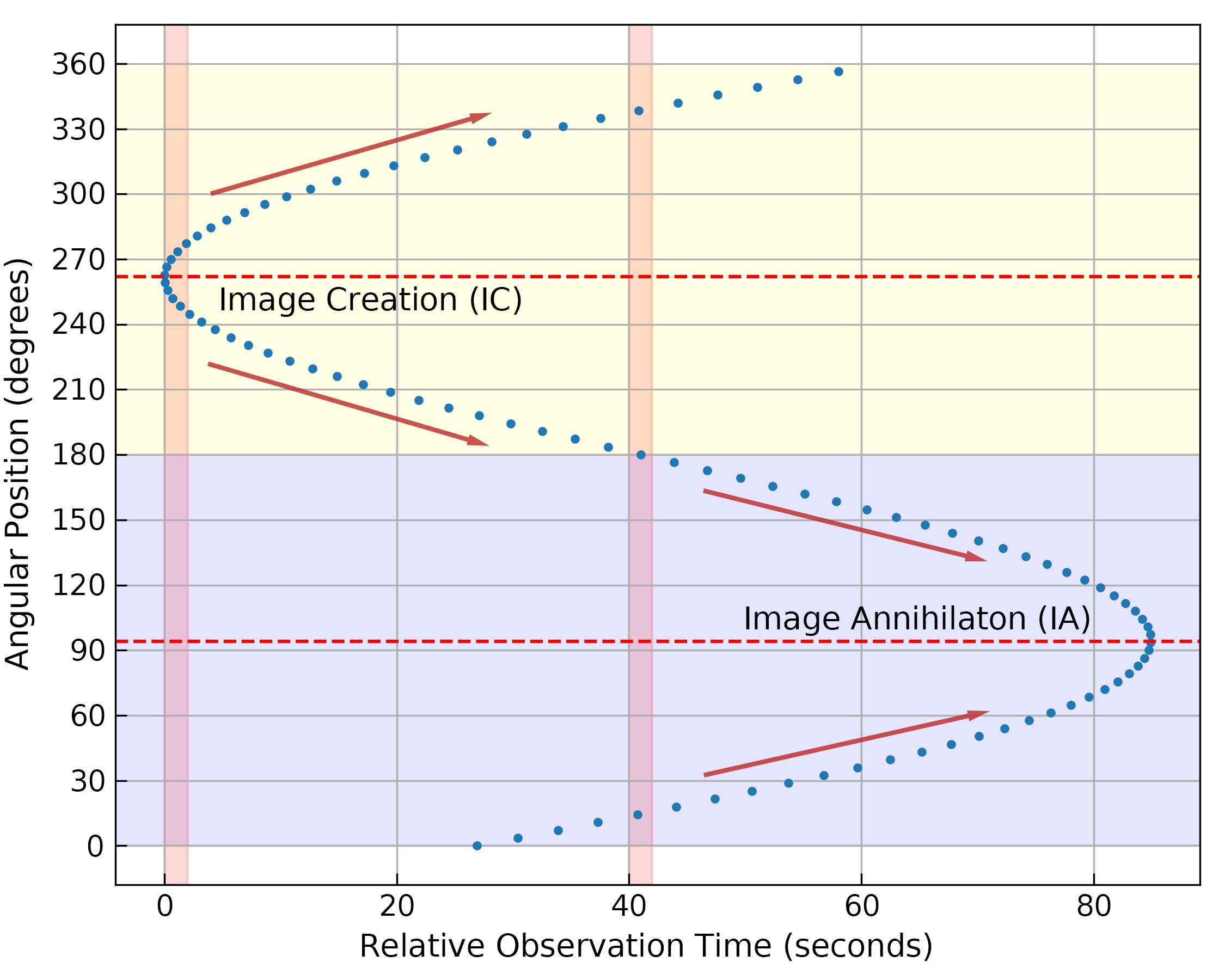}
    \centering
    \caption{An example scenario where the observation starts with an image creation (IC) near $270^\circ$ and an image annihilation (IA) near $90^\circ$. The yellow and blue shaded areas represent the nearest and farthest part of the ring, respectively. The vertical red bands represent 2 seconds of exposure time. Arrows show how two images that are created in different events are annihilated together.}
    \label{fig:multiple_creation}
\end{figure}

Contrary to classical expectations, the aforementioned faster-than-light spots projected onto the Dyson ring by the sweeping pulsar beam create an observer-measured non-uniform brightening emanating from the ring, with a light curve that can be calculated using Equation \ref{eq:ttot}. To calculate this brightening, the Dyson ring is first divided into equally-spaced azimuthal sections. Then, the total elapsed time from when the beam first left the pulsar at $\beta = 0$ to when the spot was detected by the observer is calculated: $t_{tot}$, for each azimuthal section (hereafter called a point). The relation between $t_{tot}$ and $\beta$ is shown in Figure \ref{fig:multiple_creation}. An exposure time is assumed that depends on the parameters of the pulsar, a time that enables the calculation of the number of points summed per exposure. For subluminal, non-relativistic speeds, the relative observation time is a linear sum of uniformly bright angular positions, yielding an observed uniform brightness. However, for relativistic speeds, the relation between spot speed and observed brightness is nonlinear. For superluminal speeds, this non-linearity gives way to IC and IA events. As can be seen in two different red vertical bands that represent the exposure time in Figure \ref{fig:multiple_creation}, the number of sections of the Dyson ring observed during a canonical time interval is much higher around IC and IA events, indicating a bright flash at these points. 

To compute observed brightness, Lambert's cosine law is implemented for both reflection and thermal emission cases in the brightness calculation as a multiplication factor $\cos(\theta)\sin(\beta)$. It can be readily seen that when the ring is observed face-on ($\theta=90^\circ$) the observer does not see any reflection of a spot, yielding zero brightness. Conversely, when the ring is seen edge-on ($\theta = 0$), it is assumed that the beamed spot is always visible on the farthest side of the disk. Then, when $\beta=90^\circ$, the value is at its maximum and again zero on the edges $\beta=0^\circ$ and $\beta=180^\circ$.

Due to circular symmetry, IC and IA events have the same maximum brightness when observed far from the disk. However, for a relatively nearby observer, the distance between IC and IA locations differs. This results in non-symmetrical IC/IA events and different light curves. Since they are not in angularly symmetric locations, the factor arises because Lambert's cosine law scales the brightness peaks of IC and IA events differently. However, in practice, the size of the structure is very small compared to the distance to the observer. So, near the peak, IC and IA are expected to yield nearly identical brightness curves.

The duration of exposures plays a crucial role in the detection of RID events on Dyson rings. Depending on the sweeping beam spot's speed, the number of observed IC and IA events within the exposure time differs. To accurately resolve a single IC or IA event temporally, the exposure time needs to be much less than the pulsar's period. If such a short exposure time can be achieved, significantly distinguishable IA and IC flashes can be detected on the light curve. 

\begin{figure*}
    \includegraphics[width=17cm]{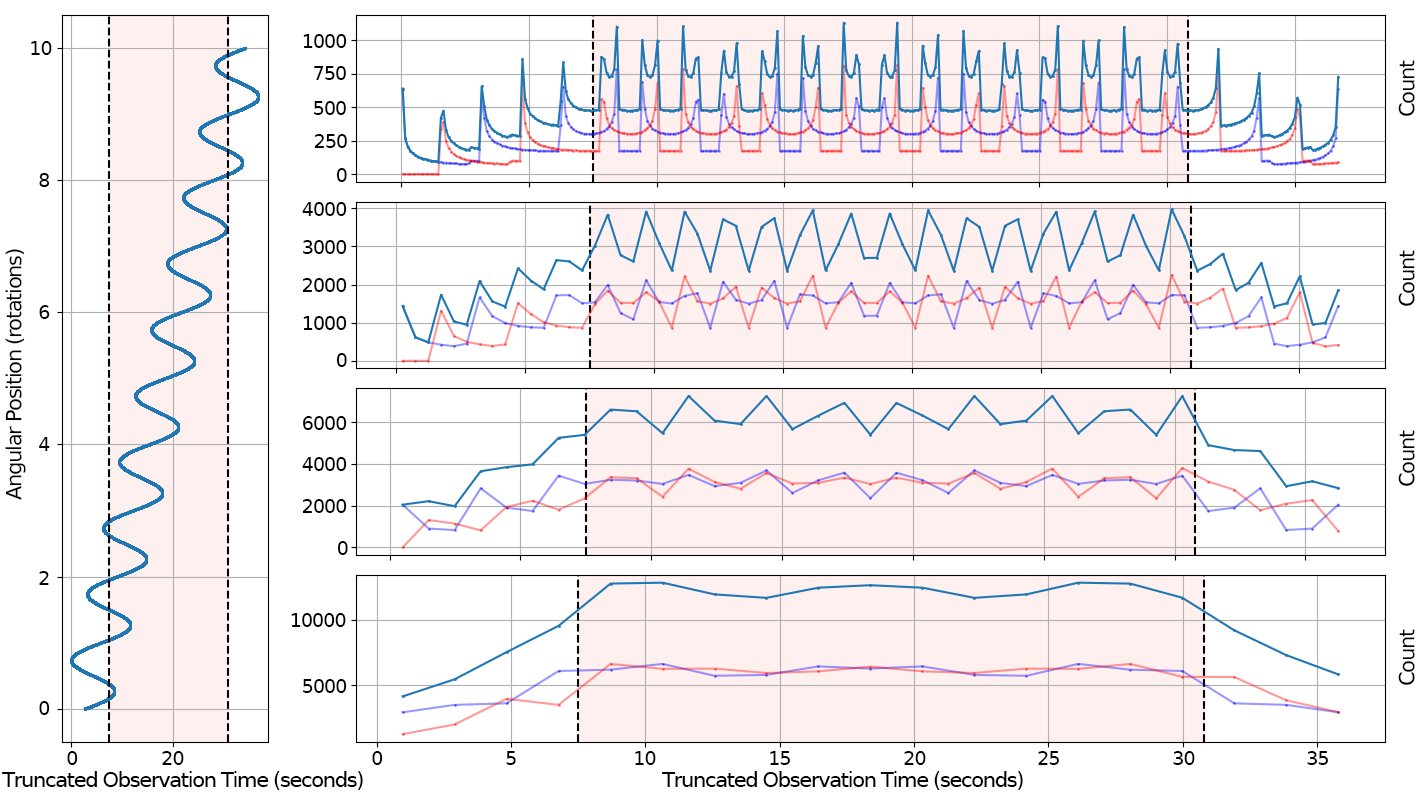}
    \centering
    \caption{Example light curves of a Dyson ring around a pulsar with a 10 $c$ beam spot speed. In the left vertical plot, the angular position over time is shown for 10 full rotations. The shaded red area represents the possible detection as the rest does not include enough binning due to initial and final conditions. On the right, four light curve plots from the top to the bottom represent 0.2 s, 0.5 s, 1 s, and 2 s exposure times, respectively. Bright red and bright blue curves on the bottom of the plots represent the two pulsar beams, while the soft blue near the top represents the binned result.}
    \label{fig:10c-20r}
\end{figure*}

Conversely, shorter exposure times may increase the difficulty of observing faint pulsars due to low SNR. Although this would limit the ability to temporally resolve a single IC or IA event, it would not limit the detection of Dyson rings. Given random phases between the pulsar's sweeping beam and the start of exposures, some exposure windows will include or exclude one extra image pair -- a windowing effect. This unpredictable windowing would cause a periodic increase and decrease in the light curve as shown in Figure \ref{fig:10c-20r}. This effect introduces a new method to detect Dyson rings around pulsars.

\begin{figure*}
    \includegraphics[width=17cm]{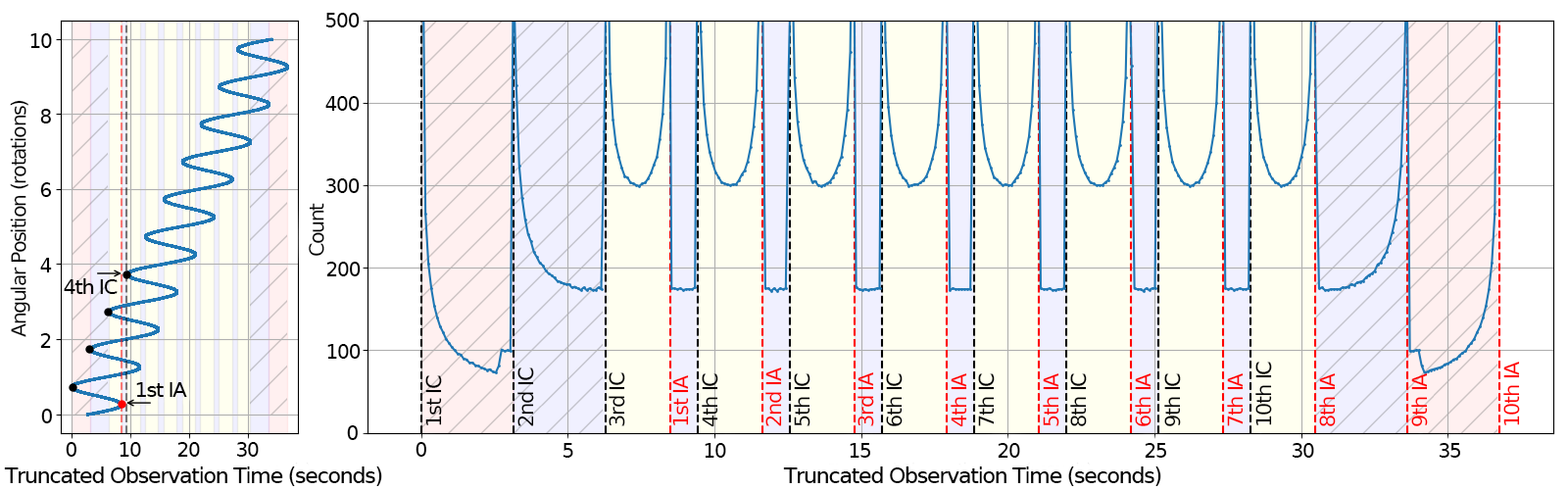}
    \centering
    \caption{An example scenario of multiple images appearing and disappearing where the beam spot's speed is 10c. The red-shaded, blue-shaded, and yellow-shaded areas correspond to 3, 5, and 7 images existing within that time interval. The unhatched area represents the actual observation as the hatched area does not include all images because of the initial and final conditions. The first IA takes place after the third IC. }
    \label{fig:number_of_images}
\end{figure*}

In cases where the beam spot sweeps around the ring much faster than the speed of light, $v \gg c$, new image pair creation events may occur before existing images undergo annihilation. This leads to the observation of multiple images of the same beam spot simultaneously. These multiple images appear through IC and disappear through IA at different times. In Figure \ref{fig:10c-20r}, a case with beam speed $v = 10 c$ and $\theta = 90^\circ$ is shown. The arbitrary exposure time represented by the red-shaded area starts with an IC, but before these two images undergo an IA, three more IC events occur. However, one of these images merges with a different already-existing image and disappears before the other. To find how many images of the beam spot would be observed at a given time, the number of intersections over a vertical line needs to be counted. In Figure \ref{fig:number_of_images}, an example vertical line is shown as a red dashed line in time, right after 1st IA, which results in an observation of five images at the same time. It is important to note that images that are formed together are never annihilated together. Therefore, although two images disappear through IA at the same point, the two images created together disappear at different times. In fact, the one image created by IC is annihilated by an image that is created on the next rotation, and the other image is annihilated by an image that is created on the previous rotation. 

For some speeds $v$, the light curve shows a nearly flat line between the peaks as shown by blue vertical bands between the yellow vertical bands in Figure \ref{fig:number_of_images}. For this certain amount of time, the ring is approximately uniformly bright. This is because after two images are annihilated, there is a time gap before the new image creation occurs. During this time gap, the pre-existing images' motion is almost linear, resulting in nearly uniform brightening. This can be seen in Figure \ref{fig:multiple_creation} where the first vertical red band shows rapidly varying speeds while the second vertical band shows almost linear motion. For different speeds $v$ this time gap can be larger or smaller.

Also, Figure \ref{fig:RID_around_circle} shows the path length of each pair that goes under annihilation. The image of the pair that moves backward has a significantly shorter path length than its forward-moving twin. This difference converges as the sweeping spot's speed gets closer to infinity where there is exactly $180^\circ$ angular separation between IC and IA events. A similar plot has been shown by \citet{Bolotovskii_1990} for the motion of a charged particle emitting radiation along a circle.

Although Figure \ref{fig:10c-20r} shows the characteristics of the light curve of Dyson rings around pulsars, because RID points are formally infinitely bright, computational inaccuracies can become significant there. One reason for this is that the number of data points within an arbitrarily defined exposure time varies, due to the phase difference between the period of the event and the exposure time (i.e. windowing effect). In addition, when the ring is divided into $N$ points to calculate $t_{tot}$ for each point, for different rotations, some points by chance fall very close to the RID point, increasing the total number of points within the exposure time. When searching for a Dyson ring around pulsars, the physical characteristics of the plot rather than the absolute values must be noted.

\section{Discussion} \label{sec:Discussion}

The rotation rate of many known pulsars, when combined with commonly assumed radii of Dyson rings, results in a faster-than-light sweeping of the pulsar's beam spot over the ring structures. In most of these scenarios, this faster-than-light sweeping results in relativistic image doubling (RID): observable image creation (IC) and image annihilation (IA) events. Depending on the period of the pulsar $p$, the radius of the ring $R$, and the inclination angle of the ring $\theta$, even multiple images of the beam spot may be created. 

In past works regarding the observability of pulsar-illuminated Dyson rings, it was assumed that the pulsar illuminated the ring uniformly \citep{osmanov_2016, Osmanov_2018, Chennamangalam_2015, Haliki_2019}. However, here it has been found that when RID effects are included, non-uniform brightening of the ring is common. Therefore, Dyson rings might have been observed previously, but not noticed.

Pulsars are not among the brightest radio sources in the sky, which makes detecting pulsar-illuminated Dyson rings challenging. When RID effects are included, however, their detection can become slightly less challenging.  This is because the onset of RID, for example, is a very bright flash which results from seeing the beamed spot along a relatively long path during only a short time. For short-duration observations, these bright RID flashes make ring detection possible at brighter survey limits, relative to non-RID ring systems. Alternatively, windowing may even make longer exposures more efficient in detecting Dyson rings than shorter exposures.  

Although observing pulsars requires the observer to be aligned with the pulsar beam, the detection of Dyson rings does not have the same constraint. On the positive side, more pulsars will illuminate rings that beam their radiation directly toward the Earth. On the negative side, being in the pulsar beam will appear much brighter than seeing only radiation reflected from the beam.

Another concern is when a Dyson ring is at least partially opaque so that the near parts of an annular ring obscure far parts. The near-ring parts also will not reflect pulsar light to the observer. However, in these cases, it is possible that thermal radiation due to the heating of the ring can be observed from any angle. This extends the possible detection of Dyson rings further beyond the already discovered pulsars. Nevertheless, given the intrinsic function of these structures in energy harvesting, the resulting radiation signatures are anticipated to be significantly smaller than the pulsars themselves. 

Although surely starting from raw ice and rock, what composite material Dyson rings are composed of can only be guessed. This composite structure could absorb pulsar radiation very efficiently, reflecting and radiating only a low percentage of incident radio power. However, some of the incident power might be radiated outside the radio band. Therefore, Dyson ring searches should not be limited to constant emission in pulsar's spectrum only, but rather focus on finding RID signatures over the entire spectrum from radio waves to gamma rays. 

\citet{Eidman_1972} showed that if the pulsar's beam interacts with the surrounding spherical layer of a medium, this creates thermal emission due to the dipole interactions which may result in faster-than-light sweeping. Although he did not acknowledge image doubling, he noted: "If the mechanism of superluminous radiation is associated with pulsar radiation, one has the possibility of explaining why some pulsars have two pulses per period.". It is noted here that one can arrive at the same conclusion by introducing the RID theory. 

More generally, the implications of the theory of RID on surrounding rings are not limited to Dyson rings or Dyson swarm structures. The present analysis is relevant to any structure that reflects impacting light or emits thermal radiation due to the pulsar's beam, which includes naturally occurring dust and gas clouds. If there is a surrounding structure, the RID theory can be critically important in determining the structure's geometry. 


\section*{Acknowledgments} \label{sec:A}
OK and RJN thank Michigan Technological University for their support during the production of this work. ChatGPT-4 has been used to generate some of the code used in this work and was subsequently checked by the author.


\section*{Data Availability}
 
The codes that produced the results and plots in this work are available from github.com/ogetaykayali.



\bibliographystyle{mnras}


\begin{thebibliography}{}
\makeatletter
\relax
\def\mn@urlcharsother{\let\do\@makeother \do\$\do\&\do\#\do\^\do\_\do\%\do\~}
\def\mn@doi{\begingroup\mn@urlcharsother \@ifnextchar [ {\mn@doi@} {\mn@doi@[]}}
\def\mn@doi@[#1]#2{\def\@tempa{#1}\ifx\@tempa\@empty \href {http://dx.doi.org/#2} {doi:#2}\else \href {http://dx.doi.org/#2} {#1}\fi \endgroup}
\def\mn@eprint#1#2{\mn@eprint@#1:#2::\@nil}
\def\mn@eprint@arXiv#1{\href {http://arxiv.org/abs/#1} {{\tt arXiv:#1}}}
\def\mn@eprint@dblp#1{\href {http://dblp.uni-trier.de/rec/bibtex/#1.xml} {dblp:#1}}
\def\mn@eprint@#1:#2:#3:#4\@nil{\def\@tempa {#1}\def\@tempb {#2}\def\@tempc {#3}\ifx \@tempc \@empty \let \@tempc \@tempb \let \@tempb \@tempa \fi \ifx \@tempb \@empty \def\@tempb {arXiv}\fi \@ifundefined {mn@eprint@\@tempb}{\@tempb:\@tempc}{\expandafter \expandafter \csname mn@eprint@\@tempb\endcsname \expandafter{\@tempc}}}

\bibitem[\protect\citeauthoryear{Bagchi}{Bagchi}{2013}]{Bagchi_2013}
Bagchi M.,  2013, International Journal of Modern Physics D, 22, 1330021

\bibitem[\protect\citeauthoryear{Baune}{Baune}{2009}]{Baune_2009}
Baune S.,  2009, \mn@doi [Physics Education] {10.1088/0031-9120/44/3/010}, 44, 296

\bibitem[\protect\citeauthoryear{Bolotovskiĭ \& Bykov}{Bolotovskiĭ \& Bykov}{1990}]{Bolotovskii_1990}
Bolotovskiĭ B.~M.,  Bykov V.~P.,  1990, \mn@doi [Soviet Physics Uspekhi] {10.1070/PU1990v033n06ABEH002601}, 33, 477

\bibitem[\protect\citeauthoryear{Cavaliere, Morrison  \& Sartori}{Cavaliere et~al.}{1971}]{Cavaliere_1971}
Cavaliere A.,  Morrison P.,   Sartori L.,  1971, \mn@doi [Science] {10.1126/science.173.3996.525}, 173, 525

\bibitem[\protect\citeauthoryear{Chennamangalam, Siemion, Lorimer  \& Werthimer}{Chennamangalam et~al.}{2015}]{Chennamangalam_2015}
Chennamangalam J.,  Siemion A.~P.,  Lorimer D.,   Werthimer D.,  2015, New Astronomy, 34, 245

\bibitem[\protect\citeauthoryear{Clerici et~al.,}{Clerici et~al.}{2016}]{Clerici_2016}
Clerici M.,  et~al., 2016, Science advances, 2, e1501691

\bibitem[\protect\citeauthoryear{DeBiase}{DeBiase}{2008}]{DeBiase_2008}
DeBiase R.~L.,  2008, in 56th International Astronautical Congress of the International Astronautical Federation, the International Academy of Astronautics, and the International Institute of Space Law. pp A4--1

\bibitem[\protect\citeauthoryear{Dyson}{Dyson}{1960}]{Dyson_1960}
Dyson F.~J.,  1960, Science, 131, 1667

\bibitem[\protect\citeauthoryear{Haliki}{Haliki}{2019}]{Haliki_2019}
Haliki E.,  2019, International Journal of Astrobiology, 18, 455

\bibitem[\protect\citeauthoryear{Haliki}{Haliki}{2020}]{Haliki_2020}
Haliki E.,  2020, \mn@doi [International Journal of Astrobiology] {10.1017/S1473550420000257}, 19, 474–481

\bibitem[\protect\citeauthoryear{Harp, Richards, Shostak, Tarter, Vakoch  \& Munson}{Harp et~al.}{2016}]{Harp_2016}
Harp G.~R.,  Richards J.,  Shostak S.,  Tarter J.~C.,  Vakoch D.~A.,   Munson C.,  2016, \mn@doi [The Astrophysical Journal] {10.3847/0004-637x/825/2/155}, 825, 155

\bibitem[\protect\citeauthoryear{Hsiao et~al.,}{Hsiao et~al.}{2021}]{Hsiao_2021}
Hsiao T. Y.-Y.,  et~al., 2021, \mn@doi [Monthly Notices of the Royal Astronomical Society] {10.1093/mnras/stab1832}, 506, 1723

\bibitem[\protect\citeauthoryear{Kardashev}{Kardashev}{1964}]{Kardashev_1964}
Kardashev N.~S.,  1964, Soviet Astronomy, 8, 217

\bibitem[\protect\citeauthoryear{Kaushal \& Nemiroff}{Kaushal \& Nemiroff}{2020}]{Kaushal_2020}
Kaushal N.,  Nemiroff R.~J.,  2020, \mn@doi [The Astrophysical Journal] {10.3847/1538-4357/ab98fa}, 898, 53

\bibitem[\protect\citeauthoryear{Kayali \& Nemiroff}{Kayali \& Nemiroff}{2022}]{Kayali_2022}
Kayali O.,  Nemiroff R.,  2022, in American Astronomical Society Meeting Abstracts. pp 148--02

\bibitem[\protect\citeauthoryear{Kijak \& Gil}{Kijak \& Gil}{2003}]{Kijak_2003}
Kijak J.,  Gil J.,  2003, Astronomy \& Astrophysics, 397, 969

\bibitem[\protect\citeauthoryear{Nemiroff}{Nemiroff}{2015}]{Nemiroff_2015}
Nemiroff R.~J.,  2015, Publications of the Astronomical Society of Australia, 32, e001

\bibitem[\protect\citeauthoryear{Nemiroff}{Nemiroff}{2018}]{Nemiroff_2018}
Nemiroff R.~J.,  2018, \mn@doi [Annalen der Physik] {https://doi.org/10.1002/andp.201700333}, 530, 1700333

\bibitem[\protect\citeauthoryear{Nemiroff}{Nemiroff}{2023}]{Nemiroff_2023}
Nemiroff R.~J.,  2023, Faster Than Light: How Your Shadow Can Do It but You Can't.
Betelgeuse Press

\bibitem[\protect\citeauthoryear{Nemiroff \& Kaushal}{Nemiroff \& Kaushal}{2020}]{Nemiroff_2020}
Nemiroff R.~J.,  Kaushal N.,  2020, \mn@doi [The Astrophysical Journal] {10.3847/1538-4357/ab6440}, 889, 122

\bibitem[\protect\citeauthoryear{Osmanov}{Osmanov}{2016}]{osmanov_2016}
Osmanov Z.,  2016, \mn@doi [International Journal of Astrobiology] {10.1017/S1473550415000257}, 15, 127–132

\bibitem[\protect\citeauthoryear{Osmanov}{Osmanov}{2018}]{Osmanov_2018}
Osmanov Z.,  2018, \mn@doi [International Journal of Astrobiology] {10.1017/S1473550417000155}, 17, 112–115

\bibitem[\protect\citeauthoryear{Osmanov \& Berezhiani}{Osmanov \& Berezhiani}{2019}]{Osmanov_2019}
Osmanov Z.,  Berezhiani V.,  2019, Journal of the British Interplanetary Society, 72, 254

\bibitem[\protect\citeauthoryear{Rayleigh}{Rayleigh}{1896}]{Rayleigh_1896}
Rayleigh L.,  1896, The Theory of Sound. Mac Millan

\bibitem[\protect\citeauthoryear{Shaw, Tudor  \& Winkworth}{Shaw et~al.}{2012}]{Shaw_2012}
Shaw E.,  Tudor V.,   Winkworth D.,  2012, Physics Special Topics, 11

\bibitem[\protect\citeauthoryear{Shostak}{Shostak}{2020}]{Shostak_2020}
Shostak S.,  2020, \mn@doi [International Journal of Astrobiology] {10.1017/S1473550420000233}, 19, 456–461

\bibitem[\protect\citeauthoryear{Szary, Zhang, Melikidze, Gil  \& Xu}{Szary et~al.}{2014}]{Szary_2014}
Szary A.,  Zhang B.,  Melikidze G.~I.,  Gil J.,   Xu R.-X.,  2014, The Astrophysical Journal, 784, 59

\bibitem[\protect\citeauthoryear{Wright}{Wright}{2020}]{Wright_2020}
Wright J.,  2020, \mn@doi [Serbian Astronomical Journal] {10.2298/saj2000001w}, pp 1--18

\bibitem[\protect\citeauthoryear{Éidman}{Éidman}{1972}]{Eidman_1972}
Éidman V.~Y.,  1972, \mn@doi [Astrophysics] {10.1007/BF01041744}, 8, 359

\makeatother
\end{thebibliography}





\bsp	
\label{lastpage}
\end{document}